\documentclass[reprint,aip,amsmth,amssymb,nolongbibliography]{revtex4-2}
\usepackage{xcolor}
\usepackage[utf8]{inputenc}
\usepackage[english]{babel}
\usepackage{graphicx}
\usepackage{comment}
\usepackage{siunitx}
\usepackage{multirow}
\usepackage{makecell}
\usepackage[normalem]{ulem}
\usepackage[hidelinks,colorlinks=false,urlcolor=blue,citecolor=black]{hyperref}
\usepackage{lipsum}  

\begin{document}
\title{Phase synchronization in a sparse network of randomly connected neurons under the effect of Poissonian spike inputs}
\author{Bruno R. R. Boaretto}
\email{bruno.boaretto@unifesp.br}
\affiliation{Institute of Science and Technology, Universidade Federal de São Paulo, São José dos Campos, São Paulo, Brazil}
\author{Paulo R. Protachevicz}
\affiliation{Physics Institute, University of S\~ao Paulo, S\~ao Paulo, SP, Brazil.}
\affiliation{Institute for Complex Systems and Mathematical Biology, SUPA, University of Aberdeen, Aberdeen, United Kingdom\\}
\author{Matheus Hansen}
\affiliation{Center for Mathematics and Applications (NOVA Math), NOVA School of Science and Technology, Universidade NOVA de Lisboa, Caparica, Portugal.}
\author{Jonas Oliveira}
\affiliation{National Institute for Space Research, Sao Jose dos Campos, 12227-010, Brazil}
\author{Alexandre C. Andreani}
\affiliation{Institute of Science and Technology, Universidade Federal de São Paulo, São José dos Campos, São Paulo, Brazil}
\affiliation{Federal Institute of São Paulo, Jacareí, São Paulo, Brazil}
\author{Elbert E. N. Macau} 
\thanks{co-senior author}
\affiliation{Institute of Science and Technology, Universidade Federal de São Paulo, São José dos Campos, São Paulo, Brazil}
\begin{abstract}
This article investigates the emergence of phase synchronization in a network of randomly connected neurons by chemical synapses. The study uses the classic Hodgkin-Huxley model to simulate the neuronal dynamics under the action of a train of Poissonian spikes. In such a scenario, we observed the emergence of irregular spikes for a specific range of conductances, and also that the phase synchronization of the neurons is reached when the external current is strong enough to induce spiking activity but without overcoming the coupling current.
Conversely, if the external current assumes very high values, then an opposite effect is observed, i.e. the prevention of the network synchronization. We explain such behaviors considering different mechanisms involved in the system, such as incoherence, minimization of currents, and stochastic effects from the Poissonian spikes. Furthermore, we present some numerical simulations where the stimulation of only a fraction of neurons, for instance, can induce phase synchronization in the non-stimulated fraction of the network, besides cases in which for larger coupling values it is possible to propagate the spiking activity in the network when considering stimulation over only one neuron.
\end{abstract}

\maketitle

\begin{quotation}
The cooperative behavior of neurons and neuronal areas associated with synchronization proves to be a fundamental neural mechanism and is relevant to many cognitive processes. The brain operates in a noisy environment due to the spontaneous activity that generates random action potentials in neurons. In this scenario, neurons are submitted to a wide diversity of inputs that are provided, for example, from ion channel flux to coupling interactions and external perturbations. Hence, the effect of noise and perturbation protocols on the spiking activity of neurons is a key topic of relevance to neuroscience being the focus of several works in the last decades. This research article aims to investigate the emergence of phase synchronization in a network of randomly connected neurons under the effect of a train of Poissonian spikes. The appearance of phase synchronization is explained by analyzing the competition between internal and external currents in the network, as well as considering the Poisson inputs only in a fraction of the neuronal network. The results shed light on the emergence mechanism behind synchronous and asynchronous activities in neuronal networks under stochastic stimuli.
\end{quotation}

\section{Introduction}

The human brain is an intricate system composed of approximately $10^{11}$ neurons connected by $10^{15}$ synapses  \cite{kandel2013principles}. Understanding the relationship between the spatiotemporal activity patterns of neurons and brain functions is a primary objective of neuroscience. The complexity of the brain arises from the cooperative interaction among neurons in response to external stimuli, which leads to spontaneous activation patterns. \cite{kandel2013principles}.

In this work, we study the phase synchronization features of a sparse network of randomly connected neurons under the effect of a train of Poissonian spikes. These types of spike inputs are thought to play an important role in generating the highly irregular spiking patterns observed in cortical neurons \cite{softky1993highly,schneidman2003synergy}. There are several lines of evidence that support the use of Poissonian spike inputs in cortical neurons \cite{softky1993highly,brunel1999fast,shadlen1998variable,stevens1998input,schneidman2003synergy}. One of the key pieces of evidence comes from studies of the statistics of natural stimuli, such as images or sounds \cite{mazzoni2008encoding}. These studies have shown that the statistical properties of natural stimuli are well-described by Poisson processes, suggesting that the brain may have evolved to process information in a way that is optimized for these statistics \cite{shadlen1998variable,renart2010asynchronous,litwin2012slow}. 

To simulate the neuronal dynamics, we use the classic Hodgkin-Huxley model \cite{hodgkin1952quantitative}, which mimics the action potential when the neuron is stimulated above a threshold \cite{izhikevich2007dynamical}. The model exhibits Hopf bifurcations as the constant inputted current is varied \cite{keener1998mathematical,izhikevich2007dynamical}, in which for a range of currents there is a stable limit cycle that gives rise to periodic spiking behavior \cite{ermentrout2010mathematical}. We show that the behavior induced by Poissonian spikes consists of irregular spikes for a specific range of conductances. As the main result, we identified the appearance of high firing frequency and synchronization in the network considering different fractions of Poissonian perturbed neurons. Our findings highlight the influence of stochastic external stimuli (Poissonian) and internal neuronal interactions (coupling) on the brain's emergence of complex firing patterns.

The neuronal activity characterized by the action potential occurs due to a process of depolarization followed by repolarization when neurons are sufficiently stimulated \cite{izhikevich2007dynamical}. When two or more neurons start their depolarization process together, the behavior can be attributed to the collective phenomenon that is associated with the more general framework of phase synchronization of oscillators \cite{ivanchenko2004phase}. All the behavioral disorders that characterize psychiatric illness (unhealthy neural behaviors) are disturbances in brain functioning \cite{kandel2013principles} and abnormal levels of synchronization have been related to unhealthy neural behaviors like epilepsy and Parkinson's disease \cite{kandel2013principles,mormann2000mean,hammond2007pathological,popovych2014control}.

The main goal of our work is to investigate how synchronization emerges in a network of randomly connected neurons of chemical synapses. We show that the network reaches phase synchronization in regimes where the external current is sufficient to induce spiking activity in the network but not overcome the coupling current.
On the other hand, greater values of external current prevent the network to synchronize due to two distinct mechanisms: stochasticity due to the randomness of the external spikes and the minimization since the external current suppresses the amplitude of the presynaptic neurons. At last, we show that stimulating a fraction of neurons can induce phase synchronization in the non-stimulated fraction while the stimulated fraction remains incoherent. Moreover, if the external current is increased, the coupling factor is minimized, losing influence in the non-stimulated fraction of the network. Furthermore, stimulating only one neuron can propagate spiking activity in the network for larger coupling values. 

This paper is organized as follows: Section \ref{sec:model} presents the neuronal model and the equations which rule the external synaptic current, Section \ref{sec:network} presents the network setup and how the phase synchronization is evaluated, the results are depicted in Section \ref{sec:results}, and Section \ref{sec:conclusions} presents the discussion and our conclusions. 

\section{Neuronal model}\label{sec:model}

To simulate the spiking neuronal dynamics, we  consider the Hodgkin-Huxley (HH) model \cite{hodgkin1952quantitative}, which was the first to describe mathematically a regenerative current that generates an action potential. The time evolution of the membrane potential of the neuron $V(t)$ measured in \SI{}{\milli\volt} (millivolts) is related to the variations of two voltage-gated channels associated with the ion concentrations of potassium ($\mathrm K ^+$) and sodium ($\mathrm{Na}^+$), as well as a leakage channel associated with the passive variations (non-gated channels) \cite{ermentrout2010mathematical}.  The time evolution of the membrane potential of the neuron $V(t)$ is given by
\begin{eqnarray}
C_\mathrm M \frac{dV}{dt}&=& -g_\mathrm K n^4(V-E_\mathrm K) - g_\mathrm{Na}m^3h(V-E_\mathrm{Na}) \nonumber \\ && - g_\mathrm{\ell}(V-E_\mathrm \ell) +  I_{\rm ext}(t), \\
\frac{dn}{dt} &=& \alpha_n(1-n) - \beta_n n, \label{eq:n}\\
\frac{dm}{dt} &=& \alpha_m(1-m) - \beta_m m,\\
\frac{dh}{dt} &=& \alpha_h(1-h) - \beta_h h, \label{eq:h}
\end{eqnarray}
where $C_\mathrm M$ is the capacitance of the cell membrane and $I_{\rm ext}$ is the external current. The parameters $g_\mathrm K$, $g_\mathrm{Na}$ and $g_\mathrm \ell$ are the maximum conductances, and $E_\mathrm K$, $E_\mathrm{Na}$, and $E_\mathrm \ell$ are the reversal potential of each ionic current. The variables $n$ and $m$ are related to the activation of the potassium and sodium ionic currents, respectively, and $h$ is the inactivation of the sodium current. $\alpha$ and $\beta$ are functions dependent on $v=V/{\rm mV}$ described as
\begin{eqnarray}\label{eq:alpha_1}
\alpha_n&=&\frac{0.01(v+55)}{(1-\exp[-(v+55)/10])},\\
\alpha_m&=&\frac{0.1(v+40)}{(1-\exp[-(v+40)/10])},\\
\alpha_h&=&0.07\exp[-(v+65)/20],\\
\beta_n&=&0.125\exp[-(v+65)/80],\\
\beta_m&=&4\exp[-(v+65)/18],\\
\beta_h&=&\frac{1}{(1+\exp[-(v+35)/10])}.\label{eq:alpha_2}
\end{eqnarray} 

\begin{figure}[htb!]
    \centering
    \includegraphics[width=.9\columnwidth]{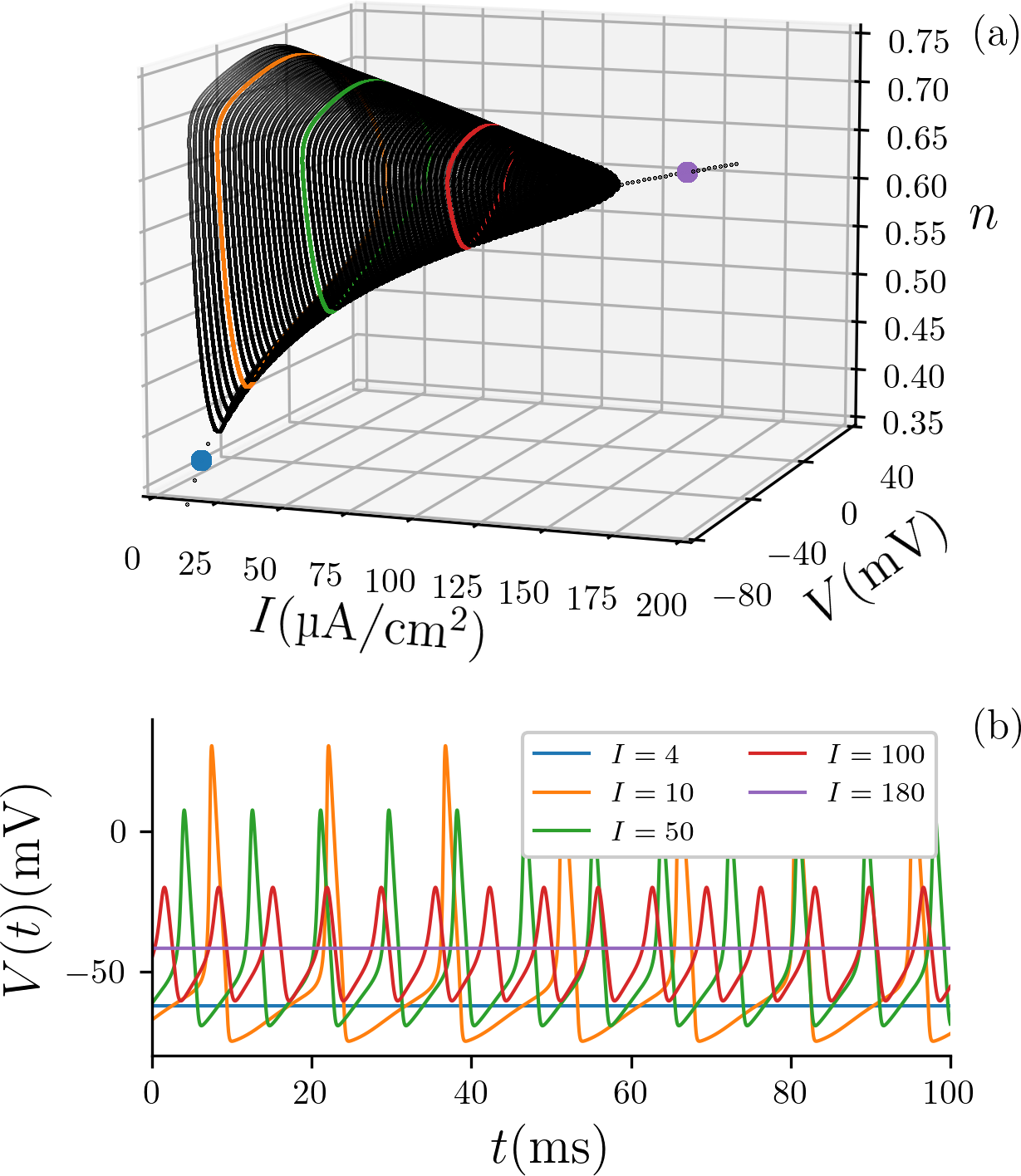}
    \caption{Dynamics of HH model under external constant current. (a) Two-dimensional projection $n\times V$ of the HH model for different values of $I$. We consider the set of constant values of Table \ref{tab:tablehh} and the initial condition $V(0)=\SI{-70}{\milli\volt}$ and $n(0)=m(0)=h(0)=0$. A transient of $\SI{1}{\second}$ was discarded. (b) Time evolution of the membrane potential $V(t)$ for colored cases.} 
    \label{fig:hh_dyn}
\end{figure}

Figure \ref{fig:hh_dyn} presents the evolution of neuronal membrane in the model as a function of a constant external current $I_\mathrm{ext}(t) = I$. The parameter $I$ is a free parameter in the model and is measured in $\SI{}{\micro\ampere/\centi\meter^2}$. Table \ref{tab:tablehh} shows the set of constant values considered in the simulation based on Ref. \cite{ermentrout2010mathematical}. Figure \ref{fig:hh_dyn} (a) depicts the two-dimensional projection $n \times V$ of the system phase portrait as a function of $I$. Figure \ref{fig:hh_dyn} (b) depicts the time evolution of $V(t)$ for colored cases shown in Figure \ref{fig:hh_dyn} (a). The colors identify the membrane evolution submitted to $I$ = 4 (blue), $I$ = 10 (orange), $I$ = 50 (green), $I$ = 100 (red), and $I$ = 180 (purple). As can be seen in the figures, constant values of membrane potential are observed for $I$ = 4 (blue) and $I$ = 180 (purple), while for the other values of external current, the membrane potential changes over time.

Considering $I$ as a bifurcation parameter, the HH model is a classic dynamical system that undergoes Hopf bifurcations \cite{izhikevich2007dynamical}. For small values of $I$, the system evolves to a stable equilibrium point (blue line). As $I$ is increased $I^* \approx 10$ the equilibrium point loses stability and gives rise to a stable limit cycle attractor due to a subcritical Andronov-Hopf bifurcation\cite{ermentrout2010mathematical}, the limit cycle characterizes the periodic orbits of the spiking activity (orange, green and red lines). The transition from the equilibrium state to the oscillatory state depends on the initial conditions for $I\approx I^*$ \cite{andreev2019chimera,hansen2022dynamics}. As the magnitude of the injected current increases, the limit cycle is folded and the spiking dynamics collapse until the unstable equilibrium point becomes stable again (purple line) due to a supercritical Andronov-Hopf bifurcation point ($I^\dag \approx 150$) \cite{ermentrout2010mathematical}. The region which characterizes the limit cycle $I^*<I<I^\dag$ delimits the excitation block of the neuron \cite{izhikevich2007dynamical}. We observe an apparent amplitude $\times$ frequency relation in the excitation block, increasing $I$ implies an increment of the frequency but the price is paid in the decrease of the amplitude. The equations are integrated using the fourth-order Runge-Kutta method considering an integration step $\Delta t=0.01$ $\SI{}{\milli\second}$.
\begin{table}[htb!]
\setlength{\tabcolsep}{7pt}
\centering
\caption{Constants values considered in the simulation of the Hodgkin-Huxley model \cite{ermentrout2010mathematical}.}
\label{tab:tablehh}
\begin{tabular}{l l r}
    \hline \hline
     {Membrane capacitance
     ($\SI{}{\micro\farad/\centi\meter^{2}}$)} & $C_\mathrm M$ & $1$ \\ \hline
    \multirow{3}{*}{Maximum conductances
($\SI{}{\milli\siemens/\centi\meter^{2}}$)}  
&$g_\mathrm{Na}$&$120$  \\ 
&$g_\mathrm K$&$36$  \\ 
&$g_\mathrm \ell$&$0.3$\\
\hline
    \multirow{3}{*}{Resting potentials
 (\SI{}{\milli\volt})}  & $E_\mathrm{Na}$&$50$ \\  & $E_\mathrm K$&$-77$ \\   & $E_\mathrm \ell$&$-54.4$    \\
    \hline
\end{tabular}
\label{tablehh}
\end{table}

In this work, we focus on studying neuronal activity under external excitatory synaptic input due to the spontaneous activity coming from external subareas of the brain \cite{ermentrout2008reliability}. These synapses are activated by random Poisson spike trains that reach the neuron with a constant rate $\nu_\mathrm{ext}$. The external synaptic current is the sum of the chemical excitatory signals given by
\begin{equation}
    I_\mathrm{ext}(t) = g_\mathrm{ext} (E_\mathrm{syn} - V)  \sum_{j} s_j(t),\label{eq:ext_syn}
\end{equation}
where $g_\mathrm{ext}$ is the external synaptic conductance which is a free parameter measured in $\SI{}{\milli\siemens/\centi\meter ^2}$, $E_\mathrm{syn}$ is the reversal potential ($E_\mathrm{syn} = \SI{40}{\milli\volt}$), and $s_j$ are the presynaptic signal from the $j$-th external spike. Every time $t$ that a $j$-th presynaptic spike occurs, $s(t)$ of the postsynaptic neuron is incremented from 0 by a difference of exponential functions \cite{brunel2003determines,cavallari2014comparison,ermentrout2010mathematical}
\begin{equation}
    s(t) = \frac{\tau_0}{\tau_\mathrm d - \tau_\mathrm r}( e^{-t/\tau_\mathrm d} - e^{-t/\tau_\mathrm r})
\end{equation}
in which $\tau_0$ is a unitary constant $\SI{1}{\milli\second}$, the decay time $\tau_\mathrm d$, and the rise time $\tau_\mathrm r$  are constants of value  $2.0\; \SI{}{\milli\second}$ and $0.4 \; \SI{}{\milli\second}$, respectively. 

\begin{figure}[htb!]
    \centering
    \includegraphics[width=.85\columnwidth]{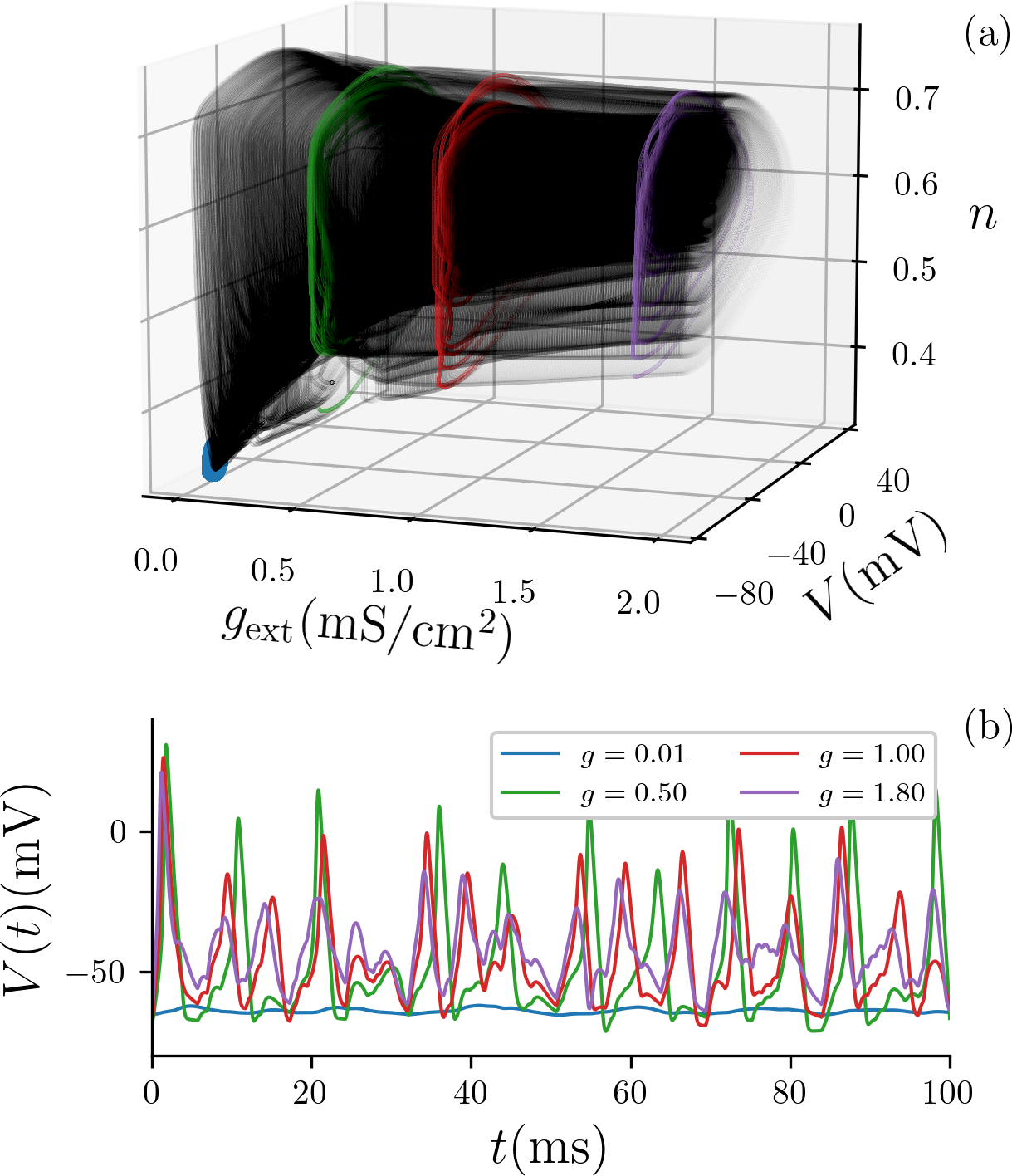}
    \caption{Dynamics of HH model under external synaptic current following a Poisson process. (a) Two-dimensional projection $n\times V$ of the system's phase portrait of the HH model for different values of $g_\mathrm{ext}$ for a fixed value of external rate $\nu=1$ $\SI{}{spike/\milli\second}$. (b) Time evolution of the membrane potential $V(t)$.}
    \label{fig:hh_dyn_poisson}
\end{figure}

Figure \ref{fig:hh_dyn_poisson} shows the dynamics of the HH model for different values of $g_\mathrm{ext}$ and fixed external rate of the Poisson process $\nu_\mathrm{ext} = 1$ \SI{}{spike/\milli\second}. Figure \ref{fig:hh_dyn_poisson} (a) depicts the two-dimensional projection $n\times V$ of HH model as a function of $g_\mathrm{ext}$. Figure \ref{fig:hh_dyn_poisson} (b) presents the time evolution of $V(t)$ for the colored cases shown in Figure \ref{fig:hh_dyn_poisson} (a). For conductance lower than $g_\mathrm{ext}=0.01$, the synaptic input is not sufficient to induce an action potential, and the membrane potential remains in a state close to the equilibrium point. Increasing the value of $g_\mathrm{ext}$, the external Poissonian current produces irregular spikes, different from the case considering a constant current where periodic spikes are generated (Fig. \ref{fig:hh_dyn}). We also observe that for greater values of $g_\mathrm{ext}$, the amplification of the synaptic current induces an increase in the spike rate and a decrease in the amplitude of the neuronal oscillation, as well as observed considering an external constant current. 

\begin{figure}[htb!]
    \centering
    \includegraphics[width=\columnwidth]{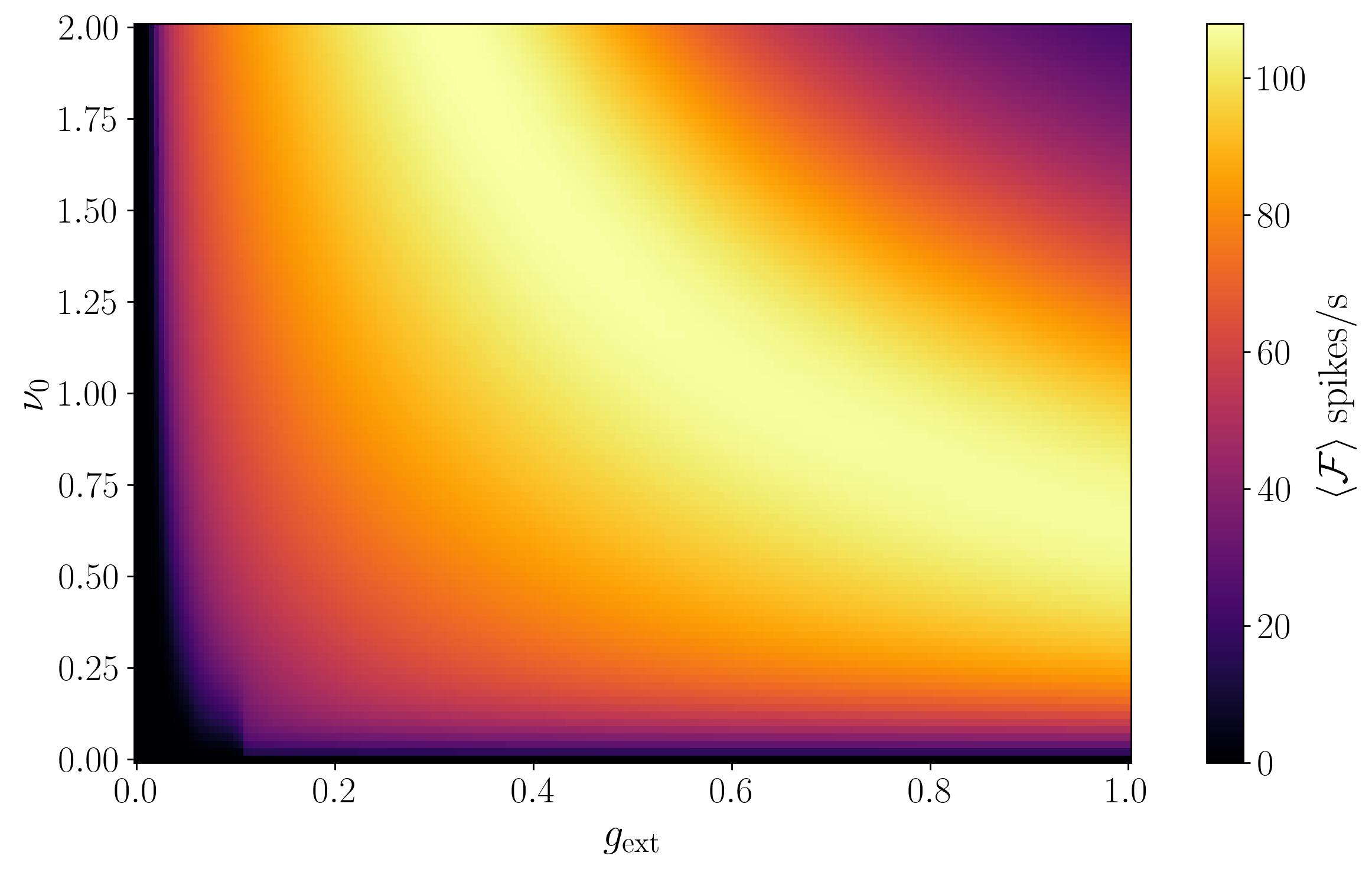}
    \caption{The average spiking rate over 100 simulations under external excitatory synaptic input  $\langle \mathcal F \rangle$.}
    \label{fig:surf_poisson}
\end{figure}

A more general framework about the spike frequency is presented in Fig. \ref{fig:surf_poisson}. Being $\mathcal F$ the number of spikes in a second per simulation, Fig. \ref{fig:surf_poisson} presents the mean value of $\mathcal F$, named $\langle \mathcal F \rangle$, which is the average over $100$ different simulations, as a function of the external conductance $g_\mathrm{ext}$ and the external rate $\nu_\mathrm{ext}$. The spike is detected when $V$ cross $-20\,\SI{}{\milli\volt}$ with a positive derivative. It is expected that increasing both conductance $g_\mathrm{ext}$ and spiking external rate $\nu_\mathrm{ext}$, the dynamical behavior transits from a steady state (black region) to an oscillatory state (colored region). In addition, it is noted that there is compensation between $\nu_\mathrm{ext}$ and $g_\mathrm{ext}$. Furthermore, higher values of $g_\mathrm{ext}$ and $\nu_\mathrm{ext}$ (a purple region located in the upper right) exhibit a decrease in $\langle\mathcal F\rangle$ corresponding to the saturation of the spiking activity illustrated in the purple line of Fig. \ref{fig:hh_dyn_poisson} and can be related with the shrink of the limit cycle which occurs for high values of external current $I$ as shown in Fig. \ref{fig:hh_dyn}. 

\section{Network setup and Synchronization quantifier}\label{sec:network}

To study the collective behavior of $N$ coupled neurons, the membrane potential of each one is  described by
\begin{eqnarray}
    C_\mathrm M \frac{dV_i}{dt}&=& -g_\mathrm K n_i^4(V_i-E_\mathrm K) - g_\mathrm{Na}m_i^3h_i(V_i-E_\mathrm{Na}) \nonumber \\ && - g_\mathrm{\ell}(V_i-E_\mathrm \ell) + I_{i,\mathrm{ext}} + I_\mathrm{i,coup},
\end{eqnarray}
in which $i$ is the neuronal index $i=1,\cdots,N$; $n_i$, $m_i$, and $h_i$ are given according to Eqs (\ref{eq:n} -- \ref{eq:h}), $I_{i,{\rm ext}}$ is the external current arriving on each neuron $i$, and $I_\mathrm{i,coup}$ is the synaptic coupled current which presents a similar form as Eq. (\ref{eq:ext_syn}), given by
\begin{equation}
    I_{i,\mathrm{coup}}(t) = \varepsilon (E_\mathrm{syn} - V_i) \sum_{j=1}^N a_{i,j} r_j(V_j),\label{eq:coup_syn}
\end{equation}
where $\varepsilon$ is the coupling parameter, $E_\mathrm{syn}$ is the reversal synaptic potential, $a_{i,j}$ is the element of the connection matrix, assuming $a_{i,j}=1$ value if there is connection from neuron $j$ to neuron $i$, otherwise $a_{i,j}=0$. The variable $r_i$ represents the fraction of bound receptors in the synapse where the kinetic model depends on the presynaptic neuron and is described by \cite{destexhe1994efficient}   
\begin{equation} 
    \frac{dr_i}{dt} = \left(\frac{1}{\tau_\mathrm{r}}-\frac{1}{\tau_\mathrm{d}}\right)\frac{1-r_i}{1+\exp[-(v_i(t)+20)]}-\frac{r_i}{\tau_\mathrm{d}}, \label{eq_r_2}
\end{equation}
in which $\tau_\mathrm{r}$ and $\tau_\mathrm{d}$ are the same parameters as defined before.

To compute phase synchronization, we use the Kuramoto order parameter \cite{kuramoto1975self}
\begin{equation}
    R = \left |\frac{1}{N}\sum_{j=1}^N e^{\imath\varphi_j(t)} \right |, \label{eq:kuramoto}
\end{equation}
where $\varphi_j$ is the phase of the $j$-th neuron, and $\imath = \sqrt{-1}$ here. In this case, $R=1$ represents a completely phase-synchronized state in which all neurons spike at the same time. Conversely, $R = 0$ means that each neuron in the network has a corresponding pair that is completely out-of-phase, this corresponds to a completely incoherent state (completely unsynchronized). In the case of a random distribution of $N$ phases, the result would be $R \sim \sqrt{1/N}$ \cite{arenas2008synchronization}. The phase of the neuron can be obtained with the relation 
\begin{equation}
\varphi_{i}(t)=2\pi k_{i} +2\pi\frac{ t-t_{{k,i}}}{t_{k+1,i}-t_{{k,i}}}, \hspace{0.5cm} t_{k,i}\leq t<t_{k+1,i},
\label{eq:phase}
\end{equation}
where $t_{k,i}$ represents the $k$-th time in which the $i$-th neuron $V$ crosses $\SI{-20}{\milli\volt}$ (spike occurrence). The phase is increased by a factor of $2\pi$ for every spike. 

\section{Results}\label{sec:results}

\begin{figure*}[htb!]
    \centering
    \includegraphics[width=2\columnwidth]{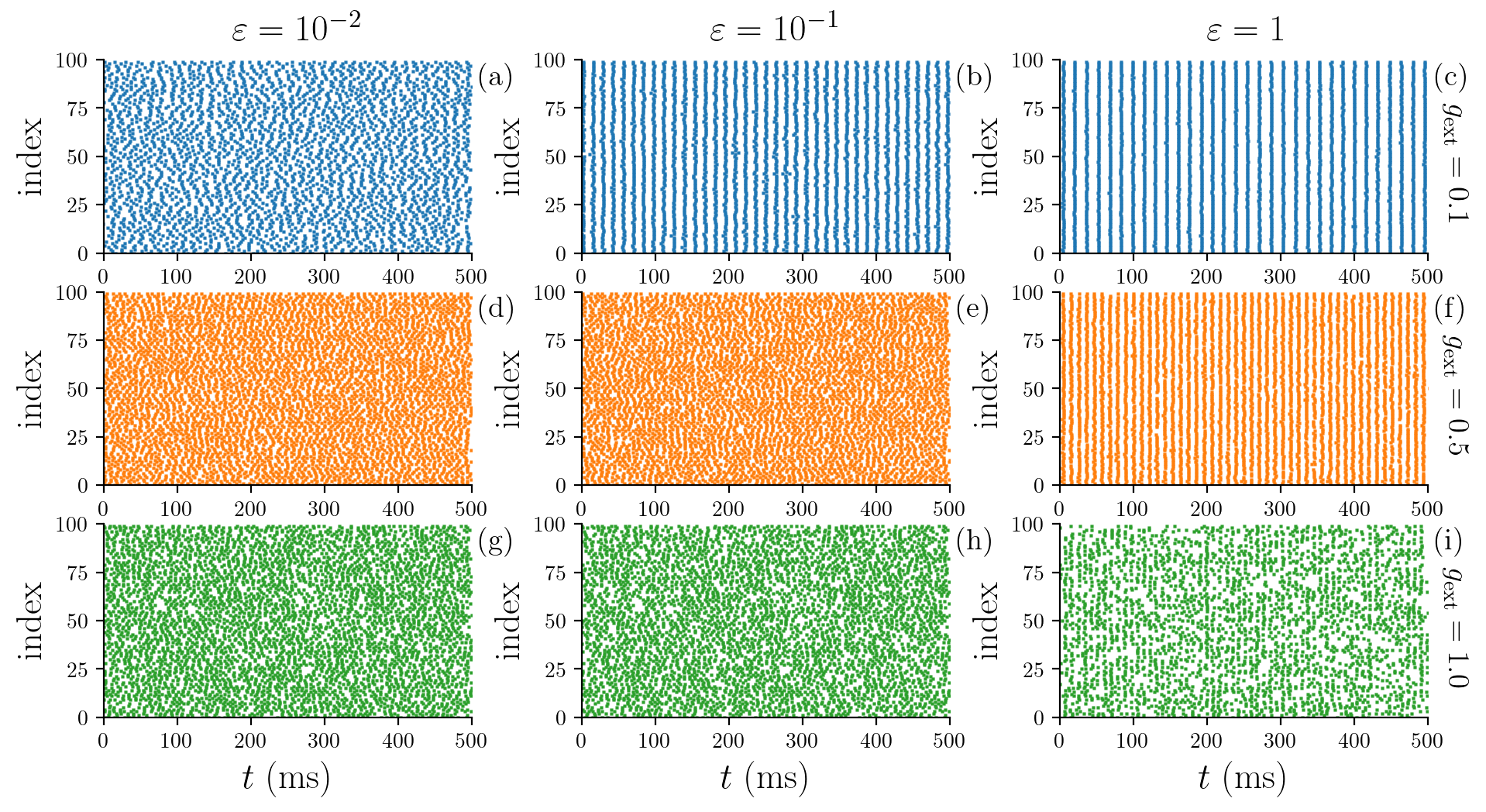}
    \caption{Temporal evolution of the network. Raster plots of the network where each dot corresponds to the beginning of a spike. The left column $\varepsilon=10^{-2}$, center column $\varepsilon=10^{-1}$, and right column $\varepsilon=1$. Each line corresponds to a fixed $g_\mathrm{ext}$, top line $g_\mathrm{ext} = 0.1$, middle line $g_\mathrm{ext} = 0.5$, and bottom line $g_\mathrm{ext} = 1.0$.} 
    \label{fig:rps}
\end{figure*}

Throughout this paper, we consider a network with $N=100$ randomly connected identical neurons, the connections follow a uniform distribution. The connection probability is fixed in $10\%$ which means that on average each neuron presents $\approx 10$ random connections. Considering the same external conductance $g_\mathrm{ext}$ and external rate $\nu_\mathrm{ext}$ for all neurons, each neuron receives its own external Poissonian train of spikes $I_{i,\mathrm{ext}}$. For simplicity, we have fixed the external rate of Poissonian spikes $\nu_\mathrm{ext}=1.0$ $\SI{}{spike/\milli\second}$ given us two free parameters: the coupling parameter $\varepsilon$, and the external conductance $g_\mathrm{ext}$. The phase synchronization is evaluated by averaging the Kuramoto order parameter on time, called mean order parameter $\langle R\rangle$, for $\SI{10}{\second}$ after discards $\SI{1}{\second}$ to avoid transient effects, a time considered sufficient to obtain the asymptotic solution of the dynamic system, and, as a result of the other quantifiers used in this work. In addition, the mean firing rate $\langle \mathcal F\rangle$ is the average over spikes produced by the network per second. The initial conditions for the neurons of the network are randomly selected from $\{V_i \in [-80,0],\; \textrm{and}\; n_i, m_i,h_i \in [0,1]\}$. To avoid any effect of the initialization in the results all the surface values are an average of over 10 different realizations considering distinct initial conditions and network configurations.

Figure \ref{fig:rps} depicts raster plots of the network where each dot corresponds to the beginning of a spike for three different values of coupling $\varepsilon = 10^{-2}$ (left column), $\varepsilon = 10^{-1}$ (center column), and $\varepsilon = 1$ (right column), and for three values of external conductance $g_\mathrm{ext} = 0.1$ (top row), $g_\mathrm{ext} = 0.5$ (middle row), and $g_\mathrm{ext} = 1.0$ (bottom row). Considering the top row, Figures \ref{fig:rps} (a) -- (c) for $g_\mathrm{ext}=0.1$, when we increase the coupling from $10^{-2}$ to $10^{-1}$, the network transits from the incoherent state to a partial phase synchronization (indicated by the vertical structures in Figure \ref{fig:rps}(b)), until the synchronized behavior for $\varepsilon=1$ (magnified in Figure \ref{fig:rps}(c)). Furthermore, comparing Figures \ref{fig:rps} (b) and (c), it is observed a decrease in spike occurrence since the number of spike trains is smaller in Figure \ref{fig:rps} (c). In the middle row, Figures \ref{fig:rps} (d) -- (f) for $g_\mathrm{ext}=0.5$, due to the magnification of the external synaptic current the transition for the synchronized state occurs only for higher values of coupling. In contrast, in the bottom row, Figures \ref{fig:rps} (g) -- (i) for $g_\mathrm{ext}=1.0$, the increase of $\varepsilon$ does not induce phase synchronization since the interplay of both external synaptic current and the coupling current saturates the spiking activity of the network.

\begin{figure}[htb!]
    \centering
    \includegraphics[width=\columnwidth]{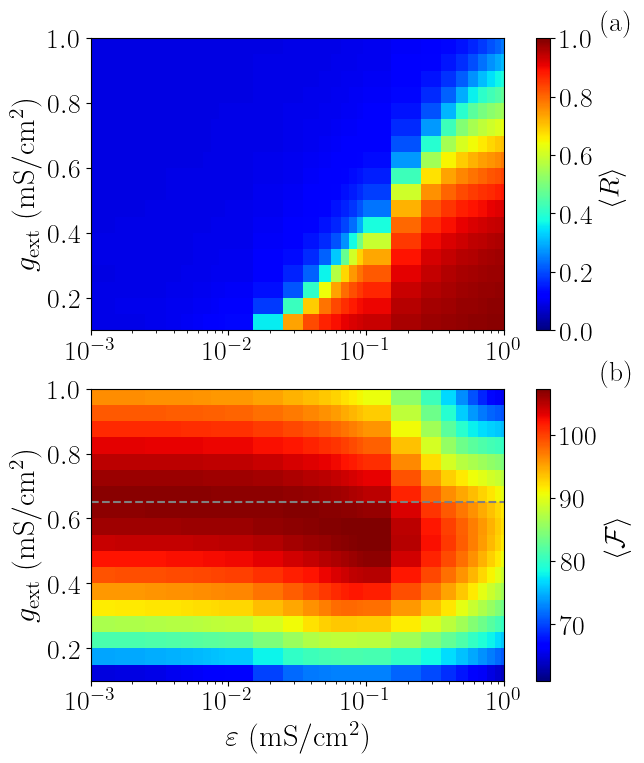}
    \caption{(a) Mean order parameter $\langle R\rangle$ and (b) mean network firing rate $\langle\mathcal F\rangle$ as a function of the coupling $\varepsilon$ and the external conductance $g_\mathrm{ext}$. The dashed line delimits the non-monotonic behavior of $\mathcal F(\varepsilon)$ from the monotonic as $\varepsilon$ increases.}
    \label{fig:surf_sync}
\end{figure}

The effect of the coupling $\varepsilon$ in association with the external conductance $g_\mathrm{ext}$ is presented in a more general scheme in Fig. \ref{fig:surf_sync}. Figure \ref{fig:surf_sync}(a) exhibits the mean order parameter ($\langle R\rangle$) while Figure \ref{fig:surf_sync}(b) shows the mean number of spikes in a second ($\langle \mathcal F\rangle$) in color codes from blue tones to red tones. The blue region in Figure \ref{fig:surf_sync}(a) exhibits low values of the order parameter $\langle R\rangle \approx 0$ that indicates an incoherent behavior among neurons of the network. As the coupling increases, there is a transition of the network to the phase synchronized regime $\langle R\rangle \approx 1$, at least for lower values of $g_\mathrm{ext} < 0.7$. For higher values of $g_\mathrm{ext}$, the stochasticity induced by the external current does not allow the network to phase synchronize. In contrast, Figure (b), $\langle \mathcal F\rangle$ depicts a non-monotonic evolution with the increase of $\varepsilon$ (below the dashed line) and a monotonic decrease (above the dashed line). This peculiarity is discussed below. 

\begin{figure}[htb!]
    \centering
    \includegraphics[width=\columnwidth]{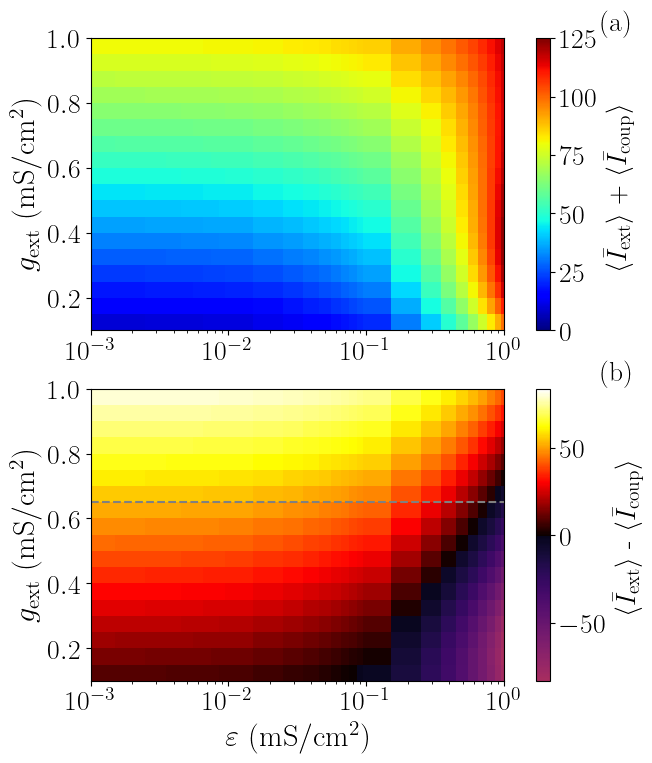}
    \caption{Interplay between the external and internal coupling current parameters. (a) Sum over contributions $\langle \bar I_\mathrm{ext} \rangle + \langle \bar I_\mathrm{coup} \rangle$ and (b) subtraction $\langle \bar I_\mathrm{ext} \rangle - \langle \bar I_\mathrm{coup} \rangle$ as a function of the coupling $\varepsilon$ and the external conductance $g_\mathrm{ext}$. The dashed line in (b) delimits the region where the $\langle \bar I_\mathrm{ext}\rangle\gtrsim \langle \bar I_\mathrm{coup}\rangle$ for the whole interval of $\varepsilon$.}
    \label{fig:surf_current}
\end{figure}

As discussed in Section \ref{sec:model}, for fixed values of Table \ref{tab:tablehh}, the activity of the neuron is determined by the current which stimulates the neuron. In this sense, in the case of coupled neurons, the excitation of the neuron depends on the interplay of the coupling current (which comes from other neurons of the network) and the external synaptic currents (which are characterized by random spikes). $\bar I_\mathrm{ext}$ and $\bar I_\mathrm{coup}$ are the mean external and coupling currents over all neurons, respectively, defined by
  \begin{eqnarray}
    \bar I_\mathrm{ext}(t) &=& \frac{1}{N}\sum_i^N I_{i,\mathrm{ext}}(t),\\
    \bar I_\mathrm{coup}(t) &=& \frac{1}{N}\sum_i^N I_{i,\mathrm{coup}}(t).
  \end{eqnarray} 
Moreover, it is possible to average these currents in time,
  \begin{eqnarray}
    \langle \bar I_\mathrm{ext}\rangle &=& \frac{1}{\tau}\sum_t^\tau \bar I_{i,\mathrm{ext}}(t),\label{eq:curr_1} \\
    \langle\bar I_\mathrm{coup}\rangle &=& \frac{1}{\tau}\sum_t^\tau \bar I_{i,\mathrm{coup}}(t).\label{eq:curr_2}
  \end{eqnarray} 
where $\tau$ corresponds to all-time instants after discards 1 s avoiding transient effects. Hence, Eqs. (\ref{eq:curr_1} - \ref{eq:curr_2}) represent the mean contribution that each current performs to the network. Figure \ref{fig:surf_current} presents in color codes in Figure \ref{fig:surf_current}(a) the sum over contributions $\langle \bar I_\mathrm{ext} \rangle + \langle \bar I_\mathrm{coup} \rangle$ and Figure (b) the subtraction $\langle \bar I_\mathrm{ext} \rangle - \langle \bar I_\mathrm{coup} \rangle$. Regards Figure \ref{fig:surf_current} (a), the total current increases with both the increment of $g_\mathrm{ext}$ and $\varepsilon$. On the other hand, in Figure \ref{fig:surf_current} (b) it is noted that $\langle \bar I_\mathrm{coup}\rangle$ gains relevance only in the purple region (lower right) which corresponds to the parameters in which the network presents a relevant phase synchronization, as can be seen in Figure \ref{fig:surf_sync} (a). In addition, the dashed line in Figure \ref{fig:surf_current} (b), delimits the region where the $\langle \bar I_\mathrm{ext}\rangle\gtrsim \langle \bar I_\mathrm{coup}\rangle$ for the whole interval of $\varepsilon$ which also delimits the boundary of the two distinct behaviors of $\langle \mathcal F\rangle$ with the increment of $\varepsilon$ observed in Figure \ref{fig:surf_sync} (b). We also noted that greater values of coupling $\varepsilon>1$ (not shown here) may lead to no spiking activity since the total current (external plus coupling) reaches high values, considerably reducing the number of spikes in the network. 

\begin{figure}[htb!]
    \centering
    \includegraphics[width=\columnwidth]{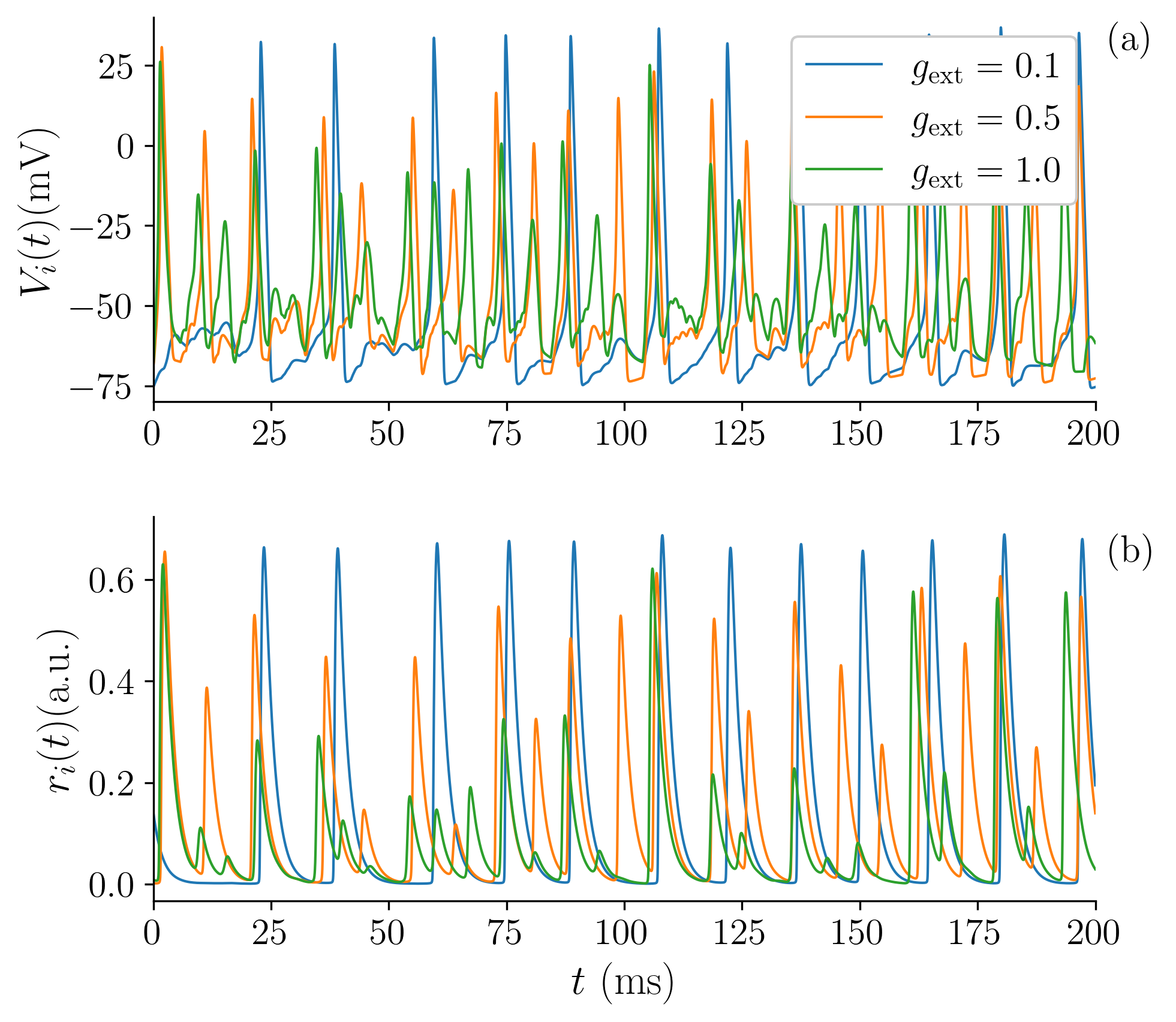}
    \caption{Temporal evolution of the membrane potential (a) and the kinetic variable (b) of an isolated neuron for different values of $g_\mathrm{ext}$. The increase in the magnitude of the external current minimizes the presynaptic effect which is propagated to the network.}
    \label{fig:poisson_synapse}
\end{figure}

The results indicate the effect of the balance between the external Poissonian signals and the internal coupling interaction of the network. Considering a null external current, $g_\mathrm{ext} = 0$ (not shown here), there is no stimulation to start the spiking activity in the network, being no longer possible to associate a phase to the neurons. For slightly greater values, like $g_\mathrm{ext}>0.1$, it is possible to start the activity in the network and the coupling can overcome the external current making a synchronized phase state possible. Conversely, at higher values of $g_\mathrm{ext} > 0.6$, the stochasticity induced by the external Poissonian signals overcomes the coupling current preventing the network from synchronizing. This happens for two particular reasons: Firstly, \textit{incoherence} since the external current is ruled by random Poissonian spikes, this irregularity disturbs the system making it hard to synchronize; Secondly, \textit{minimization of the coupling factor}, as observed in Figure \ref{fig:hh_dyn_poisson} (b), higher values of $g_\mathrm{ext}$ decrease the amplitude of the spikes, hence, the signal emitted by the presynaptic neuron, which is given by Eqs. (\ref{eq:coup_syn} - \ref{eq_r_2}), is minimized by the external current. To make the effect of minimization clearer Figure \ref{fig:poisson_synapse} (a) presents the time-evolution of the membrane potential of an isolated neuron while in Figure \ref{fig:poisson_synapse} (b) shows the kinetic variable $r_i$ (signal emitted to postsynaptic neurons). We observe that for $g_\mathrm{ext}=0.1$ (blue) the amplitude of both $V_i$ and $r_i$ are greater than the amplitude for higher values of $g_\mathrm{ext}=0.5$ (orange) and $g_\mathrm{ext}=1.0$ (green), which confirms the minimization effect of the external current on the coupling current. 

In the context of this work, different investigation lines can be considered in the research. One of the important questions is the following: What happens if only a fraction of the neurons of the network is available to receive external stimulation, and the other fraction is influenced only by coupling with these neurons? With this in mind, we separate the network into two subgroups: the first group, named Group 1 ($\Omega_1$), receives the external stimulation while the second group, named Group 2 ($\Omega_2$), $g_\mathrm{ext}=0$. This can be understood as if Group 1 shielded Group 2 from Poissonian spikes coming from the external environment.

\begin{figure*}[htb!]
    \centering
    \includegraphics[width=2\columnwidth]{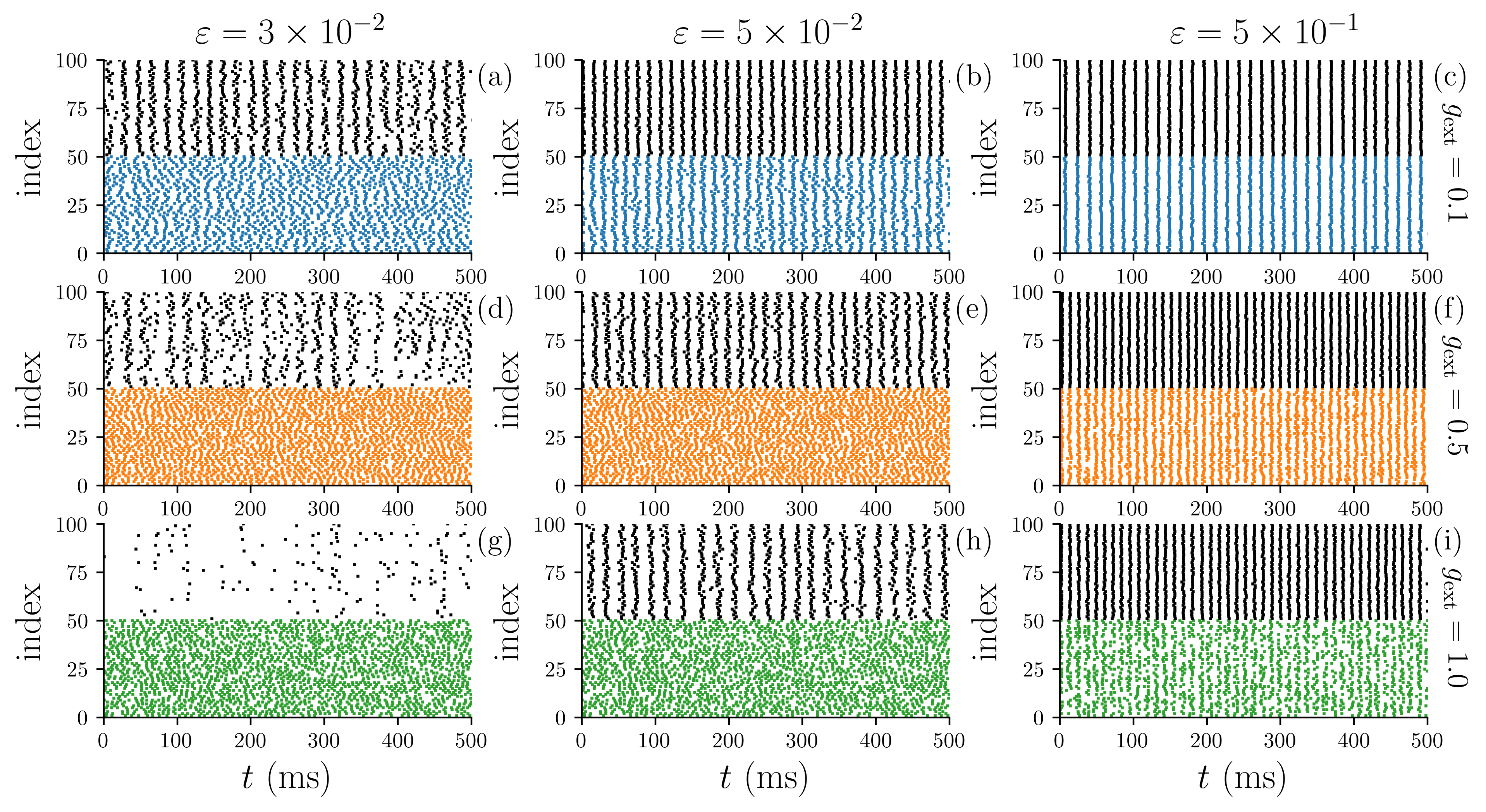}
    \caption{Temporal evolution of the two subgroups, each one with and without Poissonian external signals. Raster plots of the network where each dot corresponds to a spike. The coupling parameter is the same for all neurons $\varepsilon= 3\times 10^{-2}$ in the left column, $\varepsilon=5\times10^{-2}$ in the center column, and $\varepsilon=5\times 10^{-1}$ in the right column. We stimulate only half of the network. The stimulated neurons are represented in color codes while the non-stimulated neurons are in black ones. Each line corresponds to a fixed $g_\mathrm{ext}$, top line $g_\mathrm{ext} = 0.1$, middle line  $g_\mathrm{ext} = 0.5$, and bottom line $g_\mathrm{ext} = 1.0$.}
    \label{fig:rp_frac_0.50}
\end{figure*}

Figure \ref{fig:rp_frac_0.50} presents the raster plots of the network considering half of the network in Group 1 (colored dots), and the other half in Group 2 (black dots). We consider three different values of coupling $\varepsilon = 3\times 10^{-2}$ (left column), $\varepsilon = 5\times 10^{-2}$ (center column), and $\varepsilon = 5\times10^{-1}$ (right column), and three values of external conductance $g_\mathrm{ext} = 0.1$ (top row), $g_\mathrm{ext} = 0.5$ (middle row), and $g_\mathrm{ext} = 1.0$ (bottom row). It is observed in Figure \ref{fig:rp_frac_0.50} (a) an incoherent behavior in neurons in $\Omega_1$ while a partial phase synchronization appears in neurons in $\Omega_2$. This is an interesting phenomenon whereby the external current, which is necessary for the spiking activity in the network, overcomes the coupling current preventing phase synchronization in $\Omega_1$. However, since neurons in $\Omega_2$ are exposed only to the coupling factor, the spiking activity generated in $\Omega_1$ is sufficient to generate spikes and synchronize neurons in $\Omega_2$. Hence, in this situation, it is possible to understand that the external Poissonian spikes induce incoherence spiking activity in both $\Omega_1$ and $\Omega_2$, but in $\Omega_2$ are phase synchronized by the internal coupling. Increasing $\varepsilon$ in Figures \ref{fig:rp_frac_0.50} (b) and (c), the coupling gains relevance, and both $\Omega_1$ and $\Omega_2$ transits to phase synchronization. In these cases, the internal coupling current is strong enough to synchronize even Group 1 that are under competitive current input (external and internal).

As discussed before, there are two particular reasons why the increase in external current ($g_\mathrm{ext}$) interferes with network coupling: incoherence and minimization. In this sense, Figures \ref{fig:rp_frac_0.50} (d-f) and (g-i) (middle and bottom rows) exhibit that increasing $g_\mathrm{ext}$ makes difficult the occurrence of synchronization, producing both groups with irregular spikes (Figures \ref{fig:rp_frac_0.50} (d) and (g)). In addition, considering only the left column of Figure \ref{fig:rp_frac_0.50}, we observe that increasing $g_\mathrm{ext}$ decreases substantially the number of spikes of $\Omega_2$, the fact that emphasizing the minimization of the internal coupling current that is dependent on the membrane potential values. On the other hand, increasing $\varepsilon$, it is observed one partially synchronized group ($\Omega_2$) and one incoherent group ($\Omega_1$) (Figures (e) and (h)). In Figure  \ref{fig:rp_frac_0.50} (f) both groups are partially synchronized but $\Omega_1$ is disturbed due to the external current.  In Figure \ref{fig:rp_frac_0.50} (i), the higher external current values saturate the spiking activity in $\Omega_1$ producing incoherence while $\Omega_2$ is synchronized. We also observe that lower values of coupling can be not enough to activate the neurons in $\Omega_2$.

\begin{figure*}[htb!]
    \centering
    \includegraphics[width=2\columnwidth]{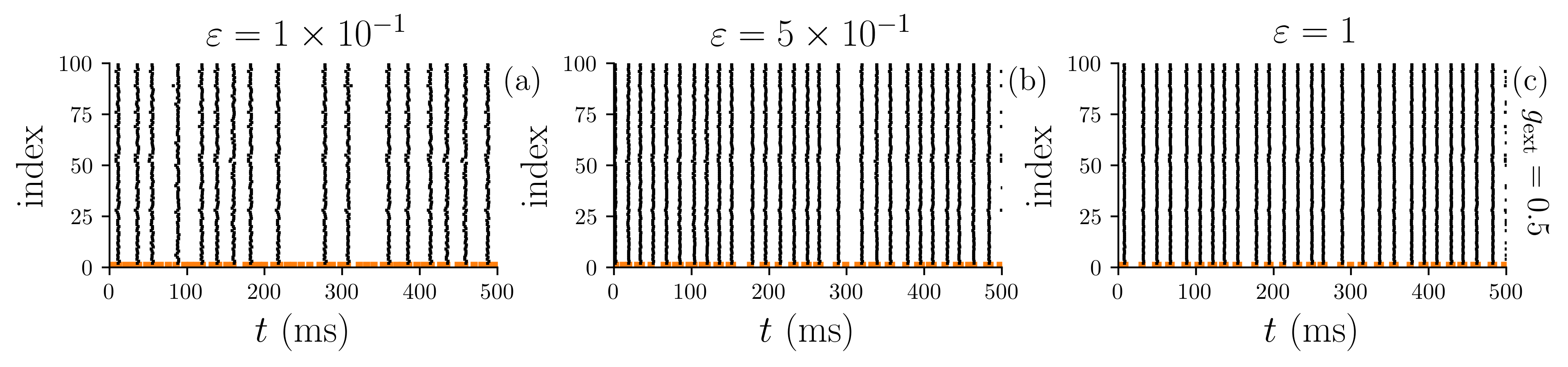}
    \caption{Temporal evolution of the network by the Poissonian excitation in only one neuron. Raster plots of the network where each dot corresponds to a spike. The coupling parameter is the same for all neurons: (a) $\varepsilon= 1\times 10{^-1}$, (b) $\varepsilon=5\times10^{-1}$, and (c) $\varepsilon=1$. We stimulate only one neuron which is represented in orange color while the non-stimulated neurons are represented in black. In this figure $g_\mathrm{ext}$ is fixed in $0.5$.}
    \label{fig:RP_frac_0.01}
\end{figure*}

A more extreme scenario is explored in Figure \ref{fig:RP_frac_0.01}, where we extrapolate the previous analysis by the excitation of only one neuron. For lower levels of coupling, $\varepsilon<0.1$ (not shown), the neuron in $\Omega_1$ spikes irregularly alone. For values greater than $\varepsilon=0.1$, the coupling is high enough to produce spiking activity and sufficient to induce phase synchronization in $\Omega_2$, as shown in Figure \ref{fig:RP_frac_0.01} (a). Increasing the coupling to $\varepsilon=0.5$, as shown in Figure \ref{fig:RP_frac_0.01} (b), increases the number of spikes in $\Omega_2$. The synchronization between $\Omega_1$ and $\Omega_2$ is greater in  Figure \ref{fig:RP_frac_0.01} (b) when compared with Figure \ref{fig:RP_frac_0.01} (a), and even magnified in Figure \ref{fig:RP_frac_0.01} (c) $\varepsilon=1$. 

 %
%
%

\section{Conclusions}\label{sec:conclusions}

Throughout this paper, we have analyzed the phase synchronization behavior of a network composed of 100 Hodgkin-Huxley neurons randomly coupled  and submitted to external Poissonian signals. In this sense, when the coupling is turned off, we show that there is a range of values in the external conductance that produces irregular spikes. Out of this range, there is no spiking activity: since for smaller values the external current is not enough to stimulate the action potential of the neuron and greater values saturate the membrane potential preventing the neuron from spiking.  

When the coupling is turned on, we take into account the interplay between the external Poissonian signals and the synaptic coupling currents. The process of phase synchronization (or partial phase synchronization) occurs when the coupling current overcomes the external current which happens for a small external conductance and great coupling conductance. In this model, the increase in external conductance disrupts the synchronization in two different ways, through the irregularity of the random external spikes (incoherence) and the decrease in the amplitude of the presynaptic membrane potential (minimization) under high-intensity of stimulation. In the same way, the increase of the external current changes how the mean firing rate of the network evolves with the increment of the coupling, non-monotonic for lower values, and monotonic for greater values. 

Lastly, we have analyzed the Poissonian excitation only on a fraction of the neurons in the network. In this case, we have shown that when only half of the network is stimulated, it is possible to induce phase synchronization in the non-stimulated group while the stimulated one is in an incoherent behavior. The phase synchronization of the whole network can be reached by increasing the coupling parameter. A different scenario is reached for greater values of the external conductance, where the coupling current is minimized by the reduction of potential membrane oscillations due to the external current, being possible to disrupt the synchronization even in the non-stimulated part of the network. We also studied the case where only one neuron is stimulated. In this case, for sufficient values of coupling, it is possible to generate spiking activity in the network, which due to the coupling current is accompanied by a synchronization behavior. 

\acknowledgments

B.R.R.B., E.E.N.M., and P.R.P. acknowledge the support of the São Paulo Research Foundation (FAPESP), Brazil, Proc. 2018/03211-6, 2020/04624-2,  2021/09839-0, and 2022/05153-9; and Financiadora de Estudos e Projetos (FINEP). M.H. is funded by national funds through the FCT - Fundação para a Ciência e a Tecnologia, I.P., under the scope of the projects UIDB/00297/2020 and UIDP/00297/2020 (Center for Mathematics and Applications). J.O. and A.C.A. are financed by the Coordenação de Aperfeiçoamento de Pessoal de Nível Superior - Brasil (CAPES) - Finance Code 001, Proc. 88887.603065/2021-00 and 88887.715012/2022-00.

\section*{Data Availability}
The data that support the ﬁndings of this study are available upon reasonable request from the authors.

\begin{thebibliography}{25}%
\makeatletter
\providecommand \@ifxundefined [1]{%
 \@ifx{#1\undefined}
}%
\providecommand \@ifnum [1]{%
 \ifnum #1\expandafter \@firstoftwo
 \else \expandafter \@secondoftwo
 \fi
}%
\providecommand \@ifx [1]{%
 \ifx #1\expandafter \@firstoftwo
 \else \expandafter \@secondoftwo
 \fi
}%
\providecommand \natexlab [1]{#1}%
\providecommand \enquote  [1]{``#1''}%
\providecommand \bibnamefont  [1]{#1}%
\providecommand \bibfnamefont [1]{#1}%
\providecommand \citenamefont [1]{#1}%
\providecommand \href@noop [0]{\@secondoftwo}%
\providecommand \href [0]{\begingroup \@sanitize@url \@href}%
\providecommand \@href[1]{\@@startlink{#1}\@@href}%
\providecommand \@@href[1]{\endgroup#1\@@endlink}%
\providecommand \@sanitize@url [0]{\catcode `\\12\catcode `\$12\catcode
  `\&12\catcode `\#12\catcode `\^12\catcode `\_12\catcode `\%12\relax}%
\providecommand \@@startlink[1]{}%
\providecommand \@@endlink[0]{}%
\providecommand \url  [0]{\begingroup\@sanitize@url \@url }%
\providecommand \@url [1]{\endgroup\@href {#1}{\urlprefix }}%
\providecommand \urlprefix  [0]{URL }%
\providecommand \Eprint [0]{\href }%
\providecommand \doibase [0]{https://doi.org/}%
\providecommand \selectlanguage [0]{\@gobble}%
\providecommand \bibinfo  [0]{\@secondoftwo}%
\providecommand \bibfield  [0]{\@secondoftwo}%
\providecommand \translation [1]{[#1]}%
\providecommand \BibitemOpen [0]{}%
\providecommand \bibitemStop [0]{}%
\providecommand \bibitemNoStop [0]{.\EOS\space}%
\providecommand \EOS [0]{\spacefactor3000\relax}%
\providecommand \BibitemShut  [1]{\csname bibitem#1\endcsname}%
\let\auto@bib@innerbib\@empty
\bibitem [{\citenamefont {Kandel}\ \emph {et~al.}(2013)\citenamefont {Kandel},
  \citenamefont {Schwartz}, \citenamefont {Jessell}, \citenamefont
  {Siegelbaum},\ and\ \citenamefont {Hudspeth}}]{kandel2013principles}%
  \BibitemOpen
  \bibfield  {author} {\bibinfo {author} {\bibfnamefont {E.~R.}\ \bibnamefont
  {Kandel}}, \bibinfo {author} {\bibfnamefont {J.~H.}\ \bibnamefont
  {Schwartz}}, \bibinfo {author} {\bibfnamefont {T.~M.}\ \bibnamefont
  {Jessell}}, \bibinfo {author} {\bibfnamefont {S.~A.}\ \bibnamefont
  {Siegelbaum}},\ and\ \bibinfo {author} {\bibfnamefont {A.~J.}\ \bibnamefont
  {Hudspeth}},\ }\href@noop {} {\emph {\bibinfo {title} {Principles of Neural
  Science}}},\ \bibinfo {edition} {5th}\ ed.\ (\bibinfo  {publisher}
  {McGraw-hill New York},\ \bibinfo {year} {2013})\BibitemShut {NoStop}%
\bibitem [{\citenamefont {Softky}\ and\ \citenamefont
  {Koch}(1993)}]{softky1993highly}%
  \BibitemOpen
  \bibfield  {author} {\bibinfo {author} {\bibfnamefont {W.~R.}\ \bibnamefont
  {Softky}}\ and\ \bibinfo {author} {\bibfnamefont {C.}~\bibnamefont {Koch}},\
  }\bibfield  {title} {\enquote {\bibinfo {title} {The highly irregular firing
  of cortical cells is inconsistent with temporal integration of random
  epsps},}\ }\href@noop {} {\bibfield  {journal} {\bibinfo  {journal} {Journal
  of Neuroscience}\ }\textbf {\bibinfo {volume} {13}},\ \bibinfo {pages}
  {334--350} (\bibinfo {year} {1993})}\BibitemShut {NoStop}%
\bibitem [{\citenamefont {Schneidman}, \citenamefont {Bialek},\ and\
  \citenamefont {Berry}(2003)}]{schneidman2003synergy}%
  \BibitemOpen
  \bibfield  {author} {\bibinfo {author} {\bibfnamefont {E.}~\bibnamefont
  {Schneidman}}, \bibinfo {author} {\bibfnamefont {W.}~\bibnamefont {Bialek}},\
  and\ \bibinfo {author} {\bibfnamefont {M.~J.}\ \bibnamefont {Berry}},\
  }\bibfield  {title} {\enquote {\bibinfo {title} {Synergy, redundancy, and
  independence in population codes},}\ }\href@noop {} {\bibfield  {journal}
  {\bibinfo  {journal} {Journal of Neuroscience}\ }\textbf {\bibinfo {volume}
  {23}},\ \bibinfo {pages} {11539--11553} (\bibinfo {year} {2003})}\BibitemShut
  {NoStop}%
\bibitem [{\citenamefont {Brunel}\ and\ \citenamefont
  {Hakim}(1999)}]{brunel1999fast}%
  \BibitemOpen
  \bibfield  {author} {\bibinfo {author} {\bibfnamefont {N.}~\bibnamefont
  {Brunel}}\ and\ \bibinfo {author} {\bibfnamefont {V.}~\bibnamefont {Hakim}},\
  }\bibfield  {title} {\enquote {\bibinfo {title} {Fast global oscillations in
  networks of integrate-and-fire neurons with low firing rates},}\ }\href@noop
  {} {\bibfield  {journal} {\bibinfo  {journal} {Neural Computation}\ }\textbf
  {\bibinfo {volume} {11}},\ \bibinfo {pages} {1621--1671} (\bibinfo {year}
  {1999})}\BibitemShut {NoStop}%
\bibitem [{\citenamefont {Shadlen}\ and\ \citenamefont
  {Newsome}(1998)}]{shadlen1998variable}%
  \BibitemOpen
  \bibfield  {author} {\bibinfo {author} {\bibfnamefont {M.~N.}\ \bibnamefont
  {Shadlen}}\ and\ \bibinfo {author} {\bibfnamefont {W.~T.}\ \bibnamefont
  {Newsome}},\ }\bibfield  {title} {\enquote {\bibinfo {title} {The variable
  discharge of cortical neurons: implications for connectivity, computation,
  and information coding},}\ }\href@noop {} {\bibfield  {journal} {\bibinfo
  {journal} {Journal of Neuroscience}\ }\textbf {\bibinfo {volume} {18}},\
  \bibinfo {pages} {3870--3896} (\bibinfo {year} {1998})}\BibitemShut {NoStop}%
\bibitem [{\citenamefont {Stevens}\ and\ \citenamefont
  {Zador}(1998)}]{stevens1998input}%
  \BibitemOpen
  \bibfield  {author} {\bibinfo {author} {\bibfnamefont {C.~F.}\ \bibnamefont
  {Stevens}}\ and\ \bibinfo {author} {\bibfnamefont {A.~M.}\ \bibnamefont
  {Zador}},\ }\bibfield  {title} {\enquote {\bibinfo {title} {Input synchrony
  and the irregular firing of cortical neurons},}\ }\href@noop {} {\bibfield
  {journal} {\bibinfo  {journal} {Nature Neuroscience}\ }\textbf {\bibinfo
  {volume} {1}},\ \bibinfo {pages} {210--217} (\bibinfo {year}
  {1998})}\BibitemShut {NoStop}%
\bibitem [{\citenamefont {Mazzoni}\ \emph {et~al.}(2008)\citenamefont
  {Mazzoni}, \citenamefont {Panzeri}, \citenamefont {Logothetis},\ and\
  \citenamefont {Brunel}}]{mazzoni2008encoding}%
  \BibitemOpen
  \bibfield  {author} {\bibinfo {author} {\bibfnamefont {A.}~\bibnamefont
  {Mazzoni}}, \bibinfo {author} {\bibfnamefont {S.}~\bibnamefont {Panzeri}},
  \bibinfo {author} {\bibfnamefont {N.~K.}\ \bibnamefont {Logothetis}},\ and\
  \bibinfo {author} {\bibfnamefont {N.}~\bibnamefont {Brunel}},\ }\bibfield
  {title} {\enquote {\bibinfo {title} {Encoding of naturalistic stimuli by
  local field potential spectra in networks of excitatory and inhibitory
  neurons},}\ }\href@noop {} {\bibfield  {journal} {\bibinfo  {journal} {PLoS
  Computational Biology}\ }\textbf {\bibinfo {volume} {4}},\ \bibinfo {pages}
  {e1000239} (\bibinfo {year} {2008})}\BibitemShut {NoStop}%
\bibitem [{\citenamefont {Renart}\ \emph {et~al.}(2010)\citenamefont {Renart},
  \citenamefont {De~La~Rocha}, \citenamefont {Bartho}, \citenamefont
  {Hollender}, \citenamefont {Parga}, \citenamefont {Reyes},\ and\
  \citenamefont {Harris}}]{renart2010asynchronous}%
  \BibitemOpen
  \bibfield  {author} {\bibinfo {author} {\bibfnamefont {A.}~\bibnamefont
  {Renart}}, \bibinfo {author} {\bibfnamefont {J.}~\bibnamefont {De~La~Rocha}},
  \bibinfo {author} {\bibfnamefont {P.}~\bibnamefont {Bartho}}, \bibinfo
  {author} {\bibfnamefont {L.}~\bibnamefont {Hollender}}, \bibinfo {author}
  {\bibfnamefont {N.}~\bibnamefont {Parga}}, \bibinfo {author} {\bibfnamefont
  {A.}~\bibnamefont {Reyes}},\ and\ \bibinfo {author} {\bibfnamefont {K.~D.}\
  \bibnamefont {Harris}},\ }\bibfield  {title} {\enquote {\bibinfo {title} {The
  asynchronous state in cortical circuits},}\ }\href@noop {} {\bibfield
  {journal} {\bibinfo  {journal} {Science}\ }\textbf {\bibinfo {volume}
  {327}},\ \bibinfo {pages} {587--590} (\bibinfo {year} {2010})}\BibitemShut
  {NoStop}%
\bibitem [{\citenamefont {Litwin-Kumar}\ and\ \citenamefont
  {Doiron}(2012)}]{litwin2012slow}%
  \BibitemOpen
  \bibfield  {author} {\bibinfo {author} {\bibfnamefont {A.}~\bibnamefont
  {Litwin-Kumar}}\ and\ \bibinfo {author} {\bibfnamefont {B.}~\bibnamefont
  {Doiron}},\ }\bibfield  {title} {\enquote {\bibinfo {title} {Slow dynamics
  and high variability in balanced cortical networks with clustered
  connections},}\ }\href@noop {} {\bibfield  {journal} {\bibinfo  {journal}
  {Nature Neuroscience}\ }\textbf {\bibinfo {volume} {15}},\ \bibinfo {pages}
  {1498--1505} (\bibinfo {year} {2012})}\BibitemShut {NoStop}%
\bibitem [{\citenamefont {Hodgkin}\ and\ \citenamefont
  {Huxley}(1952)}]{hodgkin1952quantitative}%
  \BibitemOpen
  \bibfield  {author} {\bibinfo {author} {\bibfnamefont {A.~L.}\ \bibnamefont
  {Hodgkin}}\ and\ \bibinfo {author} {\bibfnamefont {A.~F.}\ \bibnamefont
  {Huxley}},\ }\bibfield  {title} {\enquote {\bibinfo {title} {A quantitative
  description of membrane current and its application to conduction and
  excitation in nerve},}\ }\href@noop {} {\bibfield  {journal} {\bibinfo
  {journal} {The Journal of Physiology}\ }\textbf {\bibinfo {volume} {117}},\
  \bibinfo {pages} {500} (\bibinfo {year} {1952})}\BibitemShut {NoStop}%
\bibitem [{\citenamefont {Izhikevich}(2007)}]{izhikevich2007dynamical}%
  \BibitemOpen
  \bibfield  {author} {\bibinfo {author} {\bibfnamefont {E.~M.}\ \bibnamefont
  {Izhikevich}},\ }\href@noop {} {\emph {\bibinfo {title} {Dynamical systems in
  neuroscience}}}\ (\bibinfo  {publisher} {MIT press},\ \bibinfo {year}
  {2007})\BibitemShut {NoStop}%
\bibitem [{\citenamefont {Keener}\ and\ \citenamefont
  {Sneyd}(1998)}]{keener1998mathematical}%
  \BibitemOpen
  \bibfield  {author} {\bibinfo {author} {\bibfnamefont {J.}~\bibnamefont
  {Keener}}\ and\ \bibinfo {author} {\bibfnamefont {J.}~\bibnamefont {Sneyd}},\
  }\href@noop {} {\emph {\bibinfo {title} {Mathematical Physiology}}}\
  (\bibinfo  {publisher} {Springer-Verlag},\ \bibinfo {address} {New York},\
  \bibinfo {year} {1998})\BibitemShut {NoStop}%
\bibitem [{\citenamefont {Ermentrout}\ and\ \citenamefont
  {Terman}(2010)}]{ermentrout2010mathematical}%
  \BibitemOpen
  \bibfield  {author} {\bibinfo {author} {\bibfnamefont {B.}~\bibnamefont
  {Ermentrout}}\ and\ \bibinfo {author} {\bibfnamefont {D.~H.}\ \bibnamefont
  {Terman}},\ }\href@noop {} {\emph {\bibinfo {title} {Mathematical foundations
  of neuroscience}}},\ Vol.~\bibinfo {volume} {35}\ (\bibinfo  {publisher}
  {Springer},\ \bibinfo {year} {2010})\BibitemShut {NoStop}%
\bibitem [{\citenamefont {Ivanchenko}\ \emph {et~al.}(2004)\citenamefont
  {Ivanchenko}, \citenamefont {Osipov}, \citenamefont {Shalfeev},\ and\
  \citenamefont {Kurths}}]{ivanchenko2004phase}%
  \BibitemOpen
  \bibfield  {author} {\bibinfo {author} {\bibfnamefont {M.~V.}\ \bibnamefont
  {Ivanchenko}}, \bibinfo {author} {\bibfnamefont {G.~V.}\ \bibnamefont
  {Osipov}}, \bibinfo {author} {\bibfnamefont {V.~D.}\ \bibnamefont
  {Shalfeev}},\ and\ \bibinfo {author} {\bibfnamefont {J.}~\bibnamefont
  {Kurths}},\ }\bibfield  {title} {\enquote {\bibinfo {title} {Phase
  synchronization in ensembles of bursting oscillators},}\ }\href@noop {}
  {\bibfield  {journal} {\bibinfo  {journal} {Physical Review Letters}\
  }\textbf {\bibinfo {volume} {93}},\ \bibinfo {pages} {134101} (\bibinfo
  {year} {2004})}\BibitemShut {NoStop}%
\bibitem [{\citenamefont {Mormann}\ \emph {et~al.}(2000)\citenamefont
  {Mormann}, \citenamefont {Lehnertz}, \citenamefont {David},\ and\
  \citenamefont {Elger}}]{mormann2000mean}%
  \BibitemOpen
  \bibfield  {author} {\bibinfo {author} {\bibfnamefont {F.}~\bibnamefont
  {Mormann}}, \bibinfo {author} {\bibfnamefont {K.}~\bibnamefont {Lehnertz}},
  \bibinfo {author} {\bibfnamefont {P.}~\bibnamefont {David}},\ and\ \bibinfo
  {author} {\bibfnamefont {C.~E.}\ \bibnamefont {Elger}},\ }\bibfield  {title}
  {\enquote {\bibinfo {title} {Mean phase coherence as a measure for phase
  synchronization and its application to the eeg of epilepsy patients},}\
  }\href@noop {} {\bibfield  {journal} {\bibinfo  {journal} {Physica D:
  Nonlinear Phenomena}\ }\textbf {\bibinfo {volume} {144}},\ \bibinfo {pages}
  {358--369} (\bibinfo {year} {2000})}\BibitemShut {NoStop}%
\bibitem [{\citenamefont {Hammond}, \citenamefont {Bergman},\ and\
  \citenamefont {Brown}(2007)}]{hammond2007pathological}%
  \BibitemOpen
  \bibfield  {author} {\bibinfo {author} {\bibfnamefont {C.}~\bibnamefont
  {Hammond}}, \bibinfo {author} {\bibfnamefont {H.}~\bibnamefont {Bergman}},\
  and\ \bibinfo {author} {\bibfnamefont {P.}~\bibnamefont {Brown}},\ }\bibfield
   {title} {\enquote {\bibinfo {title} {Pathological synchronization in
  parkinson's disease: networks, models and treatments},}\ }\href@noop {}
  {\bibfield  {journal} {\bibinfo  {journal} {Trends in Neurosciences}\
  }\textbf {\bibinfo {volume} {30}},\ \bibinfo {pages} {357--364} (\bibinfo
  {year} {2007})}\BibitemShut {NoStop}%
\bibitem [{\citenamefont {Popovych}\ and\ \citenamefont
  {Tass}(2014)}]{popovych2014control}%
  \BibitemOpen
  \bibfield  {author} {\bibinfo {author} {\bibfnamefont {O.~V.}\ \bibnamefont
  {Popovych}}\ and\ \bibinfo {author} {\bibfnamefont {P.~A.}\ \bibnamefont
  {Tass}},\ }\bibfield  {title} {\enquote {\bibinfo {title} {Control of
  abnormal synchronization in neurological disorders},}\ }\href@noop {}
  {\bibfield  {journal} {\bibinfo  {journal} {Frontiers in Neurology}\ }\textbf
  {\bibinfo {volume} {5}},\ \bibinfo {pages} {268} (\bibinfo {year}
  {2014})}\BibitemShut {NoStop}%
\bibitem [{\citenamefont {Andreev}\ \emph {et~al.}(2019)\citenamefont
  {Andreev}, \citenamefont {Frolov}, \citenamefont {Pisarchik},\ and\
  \citenamefont {Hramov}}]{andreev2019chimera}%
  \BibitemOpen
  \bibfield  {author} {\bibinfo {author} {\bibfnamefont {A.~V.}\ \bibnamefont
  {Andreev}}, \bibinfo {author} {\bibfnamefont {N.~S.}\ \bibnamefont {Frolov}},
  \bibinfo {author} {\bibfnamefont {A.~N.}\ \bibnamefont {Pisarchik}},\ and\
  \bibinfo {author} {\bibfnamefont {A.~E.}\ \bibnamefont {Hramov}},\ }\bibfield
   {title} {\enquote {\bibinfo {title} {Chimera state in complex networks of
  bistable hodgkin-huxley neurons},}\ }\href@noop {} {\bibfield  {journal}
  {\bibinfo  {journal} {Physical Review E}\ }\textbf {\bibinfo {volume}
  {100}},\ \bibinfo {pages} {022224} (\bibinfo {year} {2019})}\BibitemShut
  {NoStop}%
\bibitem [{\citenamefont {Hansen}\ \emph {et~al.}(2022)\citenamefont {Hansen},
  \citenamefont {Protachevicz}, \citenamefont {Iarosz}, \citenamefont {Caldas},
  \citenamefont {Batista},\ and\ \citenamefont {Macau}}]{hansen2022dynamics}%
  \BibitemOpen
  \bibfield  {author} {\bibinfo {author} {\bibfnamefont {M.}~\bibnamefont
  {Hansen}}, \bibinfo {author} {\bibfnamefont {P.~R.}\ \bibnamefont
  {Protachevicz}}, \bibinfo {author} {\bibfnamefont {K.~C.}\ \bibnamefont
  {Iarosz}}, \bibinfo {author} {\bibfnamefont {I.~L.}\ \bibnamefont {Caldas}},
  \bibinfo {author} {\bibfnamefont {A.~M.}\ \bibnamefont {Batista}},\ and\
  \bibinfo {author} {\bibfnamefont {E.~E.~N.}\ \bibnamefont {Macau}},\
  }\bibfield  {title} {\enquote {\bibinfo {title} {Dynamics of uncoupled and
  coupled neurons under an external pulsed current},}\ }\href@noop {}
  {\bibfield  {journal} {\bibinfo  {journal} {Chaos, Solitons \& Fractals}\
  }\textbf {\bibinfo {volume} {155}},\ \bibinfo {pages} {111734} (\bibinfo
  {year} {2022})}\BibitemShut {NoStop}%
\bibitem [{\citenamefont {Ermentrout}, \citenamefont {Gal{\'a}n},\ and\
  \citenamefont {Urban}(2008)}]{ermentrout2008reliability}%
  \BibitemOpen
  \bibfield  {author} {\bibinfo {author} {\bibfnamefont {G.~B.}\ \bibnamefont
  {Ermentrout}}, \bibinfo {author} {\bibfnamefont {R.~F.}\ \bibnamefont
  {Gal{\'a}n}},\ and\ \bibinfo {author} {\bibfnamefont {N.~N.}\ \bibnamefont
  {Urban}},\ }\bibfield  {title} {\enquote {\bibinfo {title} {Reliability,
  synchrony and noise},}\ }\href@noop {} {\bibfield  {journal} {\bibinfo
  {journal} {Trends in Neurosciences}\ }\textbf {\bibinfo {volume} {31}},\
  \bibinfo {pages} {428--434} (\bibinfo {year} {2008})}\BibitemShut {NoStop}%
\bibitem [{\citenamefont {Brunel}\ and\ \citenamefont
  {Wang}(2003)}]{brunel2003determines}%
  \BibitemOpen
  \bibfield  {author} {\bibinfo {author} {\bibfnamefont {N.}~\bibnamefont
  {Brunel}}\ and\ \bibinfo {author} {\bibfnamefont {X.}~\bibnamefont {Wang}},\
  }\bibfield  {title} {\enquote {\bibinfo {title} {What determines the
  frequency of fast network oscillations with irregular neural discharges? i.
  synaptic dynamics and excitation-inhibition balance},}\ }\href@noop {}
  {\bibfield  {journal} {\bibinfo  {journal} {Journal of Neurophysiology}\
  }\textbf {\bibinfo {volume} {90}},\ \bibinfo {pages} {415--430} (\bibinfo
  {year} {2003})}\BibitemShut {NoStop}%
\bibitem [{\citenamefont {Cavallari}, \citenamefont {Panzeri},\ and\
  \citenamefont {Mazzoni}(2014)}]{cavallari2014comparison}%
  \BibitemOpen
  \bibfield  {author} {\bibinfo {author} {\bibfnamefont {S.}~\bibnamefont
  {Cavallari}}, \bibinfo {author} {\bibfnamefont {S.}~\bibnamefont {Panzeri}},\
  and\ \bibinfo {author} {\bibfnamefont {A.}~\bibnamefont {Mazzoni}},\
  }\bibfield  {title} {\enquote {\bibinfo {title} {Comparison of the dynamics
  of neural interactions between current-based and conductance-based
  integrate-and-fire recurrent networks},}\ }\href@noop {} {\bibfield
  {journal} {\bibinfo  {journal} {Frontiers in Neural Circuits}\ }\textbf
  {\bibinfo {volume} {8}},\ \bibinfo {pages} {12} (\bibinfo {year}
  {2014})}\BibitemShut {NoStop}%
\bibitem [{\citenamefont {Destexhe}, \citenamefont {Mainen},\ and\
  \citenamefont {Sejnowski}(1994)}]{destexhe1994efficient}%
  \BibitemOpen
  \bibfield  {author} {\bibinfo {author} {\bibfnamefont {A.}~\bibnamefont
  {Destexhe}}, \bibinfo {author} {\bibfnamefont {Z.~F.}\ \bibnamefont
  {Mainen}},\ and\ \bibinfo {author} {\bibfnamefont {T.~J.}\ \bibnamefont
  {Sejnowski}},\ }\bibfield  {title} {\enquote {\bibinfo {title} {An efficient
  method for computing synaptic conductances based on a kinetic model of
  receptor binding},}\ }\href@noop {} {\bibfield  {journal} {\bibinfo
  {journal} {Neural Computation}\ }\textbf {\bibinfo {volume} {6}},\ \bibinfo
  {pages} {14--18} (\bibinfo {year} {1994})}\BibitemShut {NoStop}%
\bibitem [{\citenamefont {Kuramoto}(1975)}]{kuramoto1975self}%
  \BibitemOpen
  \bibfield  {author} {\bibinfo {author} {\bibfnamefont {Y.}~\bibnamefont
  {Kuramoto}},\ }\bibfield  {title} {\enquote {\bibinfo {title}
  {Self-entrainment of a population of coupled non-linear oscillators},}\ }in\
  \href@noop {} {\emph {\bibinfo {booktitle} {International symposium on
  mathematical problems in theoretical physics}}}\ (\bibinfo {organization}
  {Springer},\ \bibinfo {year} {1975})\ pp.\ \bibinfo {pages}
  {420--422}\BibitemShut {NoStop}%
\bibitem [{\citenamefont {Arenas}\ \emph {et~al.}(2008)\citenamefont {Arenas},
  \citenamefont {D{\'\i}az-Guilera}, \citenamefont {Kurths}, \citenamefont
  {Moreno},\ and\ \citenamefont {Zhou}}]{arenas2008synchronization}%
  \BibitemOpen
  \bibfield  {author} {\bibinfo {author} {\bibfnamefont {A.}~\bibnamefont
  {Arenas}}, \bibinfo {author} {\bibfnamefont {A.}~\bibnamefont
  {D{\'\i}az-Guilera}}, \bibinfo {author} {\bibfnamefont {J.}~\bibnamefont
  {Kurths}}, \bibinfo {author} {\bibfnamefont {Y.}~\bibnamefont {Moreno}},\
  and\ \bibinfo {author} {\bibfnamefont {C.}~\bibnamefont {Zhou}},\ }\bibfield
  {title} {\enquote {\bibinfo {title} {Synchronization in complex networks},}\
  }\href@noop {} {\bibfield  {journal} {\bibinfo  {journal} {Physics Reports}\
  }\textbf {\bibinfo {volume} {469}},\ \bibinfo {pages} {93--153} (\bibinfo
  {year} {2008})}\BibitemShut {NoStop}%
\end{thebibliography}
%

\end{document}